\documentclass[11pt]{article}
\usepackage{amssymb}
\usepackage{epsfig}

\newcommand{\BE}{\begin{equation}}
\newcommand{\EE}{\end{equation}}
\newcommand{\BA}{\begin{eqnarray}}
\newcommand{\EA}{\end{eqnarray}}

\begin{document}



\title{Extra dimensions, preferred frames and  \\
ether-drift experiments } 

\author{
C. M. L. de Arag\~ao$^1$\thanks{Cristiane.Aragao@ct.infn.it}~, 
M. Consoli$^1$\thanks{Maurizio.Consoli@ct.infn.it}~
and A. Grillo$^2$\thanks{agrillo@dmfci.unict.it}~ \\
$^1$  {\it \small Istituto Nazionale di Fisica Nucleare, 
Sezione di Catania} \\
{\it \small  Via Santa Sofia 64, 95123 Catania, Italy } \\
\\ $^2$ {\it \small DMFCI, Facolt\`a di Ingegneria, Universit\`a di Catania} \\
{\it \small Viale Andrea Doria 6, 95126 Catania, Italy}} 
\maketitle
\thispagestyle{empty}
\date{}

\begin{abstract}
Models with extra space-time dimensions produce, tipically, a 
4D effective theory whose vacuum is not exactly Lorentz invariant but
can be considered a physical medium 
whose refractive index is determined by the gravitational field. 
This leads to a version of relativity with a preferred frame and to look
for experimental tests with the new generation of ether-drift
experiments using rotating cryogenic optical resonators.
Considering various types of cosmic
motion, we formulate precise predictions for the modulations
of the signal induced by the Earth's rotation and its orbital 
revolution around the Sun. We also compare with recent experimental
results that might represent the first modern experimental
evidence for a preferred frame.
\end{abstract}


\pagebreak

\addtocounter{page}{-1}

\section{Introduction}

Models with extra space-time dimensions \cite{extra} 
represent
an interesting approach toward a consistent quantum theory of gravity and 
its conceptual unification with the other interactions. A characteristic 
feature of such models is to
predict tipically a speed of gravity $c_g \neq c$ thus leading, in the
4D effective theory, to a version of relativity where there is 
a preferred frame $\Sigma$, 
the one associated with the isotropic value of $c_g$. 
At the same time, through the coupling to gravitons, the induced 
Lorentz-violations \cite{lorentz}
will extend to the other sectors of the theory. 
Intuitively, the effect of gravitons transforms the vacuum into
a physical medium with a 
non-trivial refractive index 
${\cal N}_{\rm vacuum} \neq 1$. 
Thus, 
if light propagates isotropically in
$\Sigma$, on the Earth there would be a small anisotropy
\BE
{{\delta c}\over{c}}\sim 
({\cal N}_{\rm vacuum}- 1)
{{v^2_{\rm earth}}\over{c^2}}
\EE
$v_{\rm earth}$ being the Earth's velocity with respect to $\Sigma$.

The aim of this paper is to explore the observable
consequences of this scenario by comparing with the ether-drift
experiments and, in particular, with the new generation where (vacuum)
cryogenic optical resonators are maintained under active rotation.
If there were a preferred frame $\Sigma$, 
one should be able to detect periodic
modulations of the signal as those associated with 
the typical angular frequency 
defined by the Earth's rotation.

To this end we shall compare with 
the results of the D\"usseldorf experiment \cite{schiller} that indeed
indicate
a definite non-zero modulation associated with the Earth's rotation and might
represent the first modern experimental evidence for a 
preferred frame. 
We shall also describe how, taking data in
different periods of the year, one can obtain precise informations to restrict
the class of possible Earth's cosmic motions.

\section{General formalism}

In general the observable implications of a speed of gravity $c_g \neq c$ 
have a considerable model dependence due to the many
possible ways of embedding in curved space-time different
graviton and photon light-cone conditions \cite{carlip}. 
Restricting to flat space, the parameter $\epsilon= c_g-1$, introduced
to parameterize the difference of the speed of gravity $c_g$ from the basic
parameter $c\equiv 1$ entering Lorentz transformations,  
is a naturally small parameter. Also, one can safely 
restrict to the case $\epsilon >0$, in view of the strong constraints placed by
the absence of gravitational Cherenkov radiation in cosmic rays \cite{nelson}.

As a convenient framework for our analysis, we shall follow the authors
of Ref.\cite{burgess} and introduce  a set
of effective Minkowski tensors $\hat{\eta}(i)_{\mu\nu}$
\BE
\hat{\eta}(i)_{\mu\nu}= \eta_{\mu\nu} - \kappa_i v_\mu v_\nu
\EE
Here
$\eta_{\mu\nu}=$diag(-1,1,1,1), $v_\mu$ is the 4-velocity of
S' with respect to a preferred frame $\Sigma$ while
$\kappa_i$ represent generalized
Fresnel's drag coefficients for particles of type $i$ 
originating from their interactions with 
the gravitons. 
In this way, the energy-momentum
relation in a given frame S' can be expressed as
\BE
          p^\mu p^\nu \hat{\eta}(i)_{\mu\nu} + m^2(i)= 0
\EE
For photons this becomes
\BE
\label{masshell}
          p^\mu p^\nu \hat{\eta}(\gamma)_{\mu\nu}= 0
\EE
with $\hat{\eta}(\gamma)_{\mu\nu}= \eta_{\mu\nu} - \kappa_\gamma v_\mu v_\nu$ 
and with a photon energy that, in the S' frame, 
depends on the direction between the photon momentum 
and the S' velocity  ${\bf{v}}$ with respect to $\Sigma$.

To obtain the photon
energy spectrum, we shall follow Jauch and Watson \cite{jauch}
who worked out the quantization of the
electromagnetic field in a moving medium. They noticed that the procedure
introduces unavoidably a preferred frame, the one where the photon 
energy does not 
depend on the direction of propagation, and which is "usually taken as
the system for which the medium is at rest". However, such an identification 
reflects the point of view of Special Relativity with no preferred frame. 
Therefore, we shall adapt their results to our case where the photon energy
does not depend on the angle in some frame $\Sigma$. In this way, in a moving
frame S', we get the radiation field Hamiltonian 
\BE
H_0=\sum_{r=1,2}\int d^3{\bf{p}}\left[\hat{n}_r({\bf{p}}) + 
{{1}\over{2}}\right]
E(|{\bf{p}}|, \theta)
\EE
where $\hat{n}_r({\bf{p}})$ is the photon number operator and
\BE
\label{watson}
E(| {\bf{p}}| , \theta)= {{ \kappa_\gamma v_0 \zeta
+ \sqrt{ |{\bf{p}}|^2(1+\kappa_\gamma v^2_0) - \kappa_\gamma \zeta^2 }}\over
{ 1 + \kappa_\gamma v^2_0}}
\EE
with
\BE
\zeta={\bf{p}}\cdot{\bf{v}}= |{\bf{p}}||{\bf{v}}|\cos\theta
\EE
$\theta\equiv\theta_{\rm lab}$ being the angle defined, in the S' frame, 
between the photon 
momentum and the S' velocity ${\bf{v}}$ with respect to $\Sigma$. 
Notice that only one of the two roots of Eq.(\ref{masshell}) appears
and the energy is not positive definite in connection with the critical
velocity $1/\sqrt{1+\kappa_\gamma}$ 
defined by the occurrence of the Cherenkov radiation.

Using the above relation,
the one-way speed of light in the S' frame depends on $\theta$ 
 (we replace
$v=|{\bf{v}}|$ and $v^2_0=1 +v^2$) 
\BE
{{E(| {\bf{p}}| , \theta)}\over{|{\bf{p}}|}}=
c_\gamma(\theta)=
{{ \kappa_\gamma v \sqrt{1+v^2} \cos\theta + 
\sqrt{ 1+ \kappa_\gamma+ \kappa_\gamma v^2 \sin^2\theta} } 
\over{1+ \kappa_\gamma(1+v^2)}}
\EE
This is different from the $v=0$ result, in the $\Sigma$ frame, where the 
energy does not depend on the angle
\BE
\label{esigma}
{{E^{(\Sigma)}(| {\bf{p}}|)}\over{|{\bf{p}}|}}=
c_\gamma= {{1}\over{ {\cal N}_{\rm vacuum} }} 
\EE
and the speed of light 
is simply rescaled by the inverse of the vacuum refractive index 
\BE
{\cal N}_{\rm vacuum}=\sqrt{1+ \kappa_\gamma}
\EE 
Working to ${\cal O}(\kappa_\gamma)$ and 
${\cal O}(v^2)$, one finds in the S' frame
\BE
\label{oneway}
c_\gamma(\theta)= {{1+ \kappa_\gamma v \cos\theta - 
{{\kappa_\gamma}\over{2}} v^2(1+\cos^2\theta) }\over
{\sqrt{ 1+\kappa_\gamma} }}
\EE
This expression differs from Eq.(6) of Ref.\cite{pla}, 
for the replacement $\cos\theta \to -\cos\theta$ 
and for the relativistic aberration 
of the angles. In Ref.\cite{pla}, in fact, the one-way speed of light in the S'
frame was parameterized in terms of the angle $\theta\equiv\theta_\Sigma$, 
between the velocity of S'
and the direction of propagation of light, as defined in the $\Sigma$ frame. 
In this way, starting from Eq.(\ref{oneway}), replacing $\cos\theta \to -
\cos\theta$ and using the aberration relation
\BE
           \cos (\theta_{\rm lab})={{-v + \cos\theta_\Sigma}\over
                                   {1-v\cos\theta_\Sigma}}
\EE
one re-obtains Eq.(6) of Ref.\cite{pla} in terms of $\theta=\theta_\Sigma$.

Finally, using Eq.(\ref{oneway}), the two-way speed of light (in terms of
$\theta=\theta_{\rm lab}$) is
\BE
\label{twoway}
\bar{c}_\gamma(\theta)= 
{{ 2  c_\gamma(\theta) c_\gamma(\pi + \theta) }\over{ 
c_\gamma(\theta) + c_\gamma(\pi + \theta) }} \sim 
1-\left[\kappa_\gamma -
{{\kappa_\gamma}\over{2}} \sin^2\theta \right]v^2
\EE
Therefore, re-introducing, for sake of clarity, 
the speed of light entering Lorentz 
transformations, $c=2.997..\cdot10^{10}$ cm/s, one can define 
the RMS \cite{robertson,mansouri} parameter 
$(1/2 -\beta +\delta)$. This is used to parameterize
the anisotropy of the speed of light
{\it in the vacuum}, through the relation 
\BE
\label{rms}
{{\bar{c}_\gamma(\pi/2 +\theta)- \bar{c}_\gamma (\theta)} \over
{\langle \bar{c}_\gamma \rangle }} \sim 
(1/2-\beta +\delta) 
{{v^2 }\over{c^2}} \cos(2\theta)
\EE
so that one can relate $\kappa_\gamma$
 to $(1/2 -\beta +\delta)$ through
\BE
\label{rmsgamma}
(1/2 -\beta +\delta)= {{\kappa_\gamma}\over{2}} 
\EE
Now, in Ref.\cite{burgess}, estimates of 
$\kappa_\gamma$ were obtained by computing the coupling of photons to
gravitons including the first few graviton loops. In this way, one obtains
typical values $\kappa_\gamma= {\cal O}(10^{-10})$ or smaller.

However, in principle, besides the graviton loops, another class of effects 
arise when considering the propagation of photons in a 
{\it background} gravitational field, such as on the Earth's surface. 
As it is well known, resumming such tree-level 
background graviton graphs leads to the realm of classical General 
Relativity where such interaction effects can be re-absorbed into a 
re-definition of the space-time metric that depends on the
external gravitational potential. However, comparing the local distortions of
space-time with the density variations of a medium, these effects can also
be incorporated into an effective refractive index. We have only to take 
into account that
$c_g\neq 1$ and that there might be a preferred frame $\Sigma$ where light
propagates isotropically.

Now, for a static gravitational field 
the first modification is trivial. 
In fact, the time-averaged scalar graviton propagator 
\BE
\lim_{T \to \infty} \langle  D({\bf{r}},t) \rangle_T =
\lim_{T \to \infty}  \int^{+T}_{-T} dt
\int  {{ d^3 {\bf{p}} }\over{(2\pi)^3}} e^{ i{\bf{p}} \cdot {\bf{r}} }
\int {{dp_0}\over{2\pi i}} 
{{ e^{-ip_0t} }\over{ c^2_g{\bf{p}}^2 -p^2_0- i\epsilon}}= 
{{1}\over{4\pi^2c^2_g r}}
\EE
is just rescaled by an overall 
factor $1/c^2_g$. This is an unobservable change where one simply
replaces the Newton constant $G^{(0)}_N$ (for $c_g=1$) with 
$G^{(0)}_N/c^2_g$ and the gravitational potential
$\varphi^{(0)}$ with $ \varphi^{(0)}/c^2_g\equiv \varphi$.

The second modification, on the other hand, requires to re-consider the 
traditional point of view on the energy of a photon in a gravitational 
field. For instance, let us consider the Earth's gravitational field 
and an observer S' placed on the Earth's surface
(but otherwise in free fall with respect to any other gravitational field). 
According to standard General Relativity, 
light is seen to propagate isotropically by S'. In fact, 
introducing the Newtonian potential
\BE
      \varphi =- {{G_N M_{\rm earth}}\over{c^2 R_{\rm earth} }} \sim
-0.7\cdot 10^{-9}
\EE
and considering the weak-field isotropic form of the metric
\cite{weinberg}
\BE
ds^2= (1+ 2\varphi) dt^2 - (1-2\varphi)(dx^2 +dy^2 +dz^2)
\EE
the energy of a photon for S' is generally assumed to be 
\BE
\label{shouldbe}
             E(|{\bf{p}}|)= c_\gamma |{\bf{p}}|
\EE
(as in Eq.(\ref{esigma})) in terms of the effective vacuum refractive index
${\cal N}_{\rm vacuum}$ in the gravitational field
\BE
                c_\gamma= {{1}\over{ 
{\cal N}_{\rm vacuum} }}\sim 1+ 2\varphi
\EE
This type of reasoning has to be modified in the presence of
a preferred frame $\Sigma$. In fact, 
it is now perfectly legitimate \cite{pagano} to ask
whether photons are seen to propagate isotropically by the S' observer
placed on the Earth's surface or by the 
$\Sigma$ observer. In the latter case, the S' energy would {\it not} 
be given by Eq.(\ref{shouldbe}) but 
would rather be given by Eq.(\ref{watson}) with a value
\BE
\label{kappagamma}
               \kappa_\gamma=
             {\cal N}^2_{\rm vacuum} - 1 \sim 28\cdot 10^{-10}
\EE
corresponding to a RMS parameter
\BE
\label{kapparms}
(1/2-\beta +\delta)\sim 
             {\cal N}_{\rm vacuum} - 1 \sim 14\cdot 10^{-10}
\EE
In this sense, as with the graviton loops considered
in Ref.\cite{burgess}, a background gravitational field transforms
the (local) vacuum into a physical medium where
the speed of light differs from the parameter $c$ entering Lorentz
transformations. If there were a preferred frame, one should detect 
an anisotropy of the two-way speed of light in modern ether-drift experiments.

\section{Cosmic motions and ether-drift experiments}

In modern ether-drift experiments, 
one measures the relative frequency shift 
$\delta \nu$ of two vacuum cryogenic optical resonators
under the Earth's rotation \cite{muller} or 
upon active rotations of the apparatus \cite{schiller}. 
If there is a preferred frame $\Sigma$, using Eqs.(\ref{twoway}) and
(\ref{rms}), the frequency shift of two orthogonal optical resonators
to ${\cal O}({{v^2}\over{c^2}})$ can be expressed as
\BE
\label{basic}
               {{\delta \nu (\theta) }\over{\nu}}= 
{{\bar{c}_\gamma(\pi/2 +\theta)- \bar{c}_\gamma (\theta)} \over
{\langle \bar{c}_\gamma \rangle }} = {{A}\over{\nu}} \cos(2\theta)
\EE
where $\theta=0$ 
indicates the direction of the ether-drift and the
amplitude of the signal is given by
\BE
\label{amplitude}
{{A}\over{\nu}}= 
(1/2 -\beta +\delta) 
{{v^2 }\over{c^2}}
\EE
$v$ denoting the
projection of the Earth's velocity with respect to $\Sigma$ 
in the plane of the interferometer. 

Notice that, in principle, one might also consider 
the possibility of measuring the frequency shift with light propagating in
a medium of refractive index 
${\cal N}_{\rm medium}\sim 1$. In this case, by continuity, very
small deviations of the refractive index from the vacuum 
value cannot qualitatively change the main result that
light propagates isotropically 
in $\Sigma$ and not in the moving
frame S' where the interferometer is at rest.
On the other hand, substantial changes 
of the refractive index, as for instance for
${\cal N}_{\rm medium}\sim 3$ which is the
relevant one for the resonating cavities of Ref.\cite{tobar}, might induce
a transition to a completely different regime where it is
the medium itself to set up the frame where light propagates 
isotropically. For this reason, in principle, different types of ether-drift
experiments might provide qualitatively different informations on the 
very existence of $\Sigma$. 

To compare with the vacuum experiment of Ref.\cite{schiller}, 
it is convenient to re-write Eq.(\ref{basic}) in the form
of Ref.\cite{schiller} where the
frequency shift at a given time $t$ is expressed as
\BE
\label{basic2}
     {{\delta \nu [\theta(t)]}\over{\nu}} = 
\hat{B}(t)\sin 2\theta(t) + 
\hat{C}(t)\cos 2\theta(t)
\EE
$\theta(t)$ being the angle of rotation of the apparatus,
  $\hat{B}(t)\equiv 2B(t)$ and $\hat{C}(t)\equiv 2C(t)$ so that
one finds an experimental amplitude
\BE
\label{expampli}
A^{\rm exp}(t)= \nu \sqrt { \hat{B}^2(t) + \hat{C}^2(t) }
\EE
Let us first consider the average signal detected in Ref.\cite{schiller}
where the relevant value is $\nu \sim 2.8\cdot 10^{14}$ Hz. 
In this case, the experimental results obtained around February 6th, 2005, 
namely
$\langle \hat{B}\nu \rangle \sim 2.8$ Hz and
$\langle \hat{C}\nu \rangle \sim -3.3$ Hz, correspond to an average amplitude
\BE
\label{expampli2}
\langle A^{\rm exp}\rangle \sim 4.3~~{\rm Hz}
\EE
This should be compared with the value of
Eq.(\ref{amplitude}), for $(1/2 -\beta +\delta)\sim 14\cdot 10^{-10}$
and a reference value $v=300$ km/s, 
$\langle A\rangle \sim 0.4$ Hz. 
Therefore, as suggested by the same authors of Ref.\cite{schiller}, 
for a meaningful comparison, we shall not consider 
the mean value (which likely
contains systematic effects of thermal origin \cite{schiller}) and
restrict the analysis to the time modulations of the signal.

In Ref.\cite{schiller}, these were parameterized as
($\omega_{\rm sid}={{2\pi}\over{23^{h}56'}}$)
\BE
\hat{C}(t)=C_0 + 
C_1\sin(\omega_{\rm sid}t)+ C_2\cos(\omega_{\rm sid}t)+
C_3\sin(2\omega_{\rm sid}t)+ C_4\cos(2\omega_{\rm sid}t)
\EE
with an analogous expression for the $\hat{B}(t)$ amplitude.
The experimental results (obtained around February 6th, 2005) can be cast into
the form
\BE
\label{sid}
           C (\omega_{\rm sid})\equiv \sqrt{C^2_1 + C^2_2}
  \sim (11 \pm 2) \cdot 10^{-16}
\EE
and
\BE
\label{2sid}
           C (2\omega_{\rm sid})\equiv \sqrt{C^2_3 + C^2_4}
  \sim (1 \pm 2) \cdot 10^{-16}
\EE
To compare with the cosmic motion defined by the CMB
it is convenient to use
the relations \cite{schiller} obtained from Ref.\cite{mewes}
\BE
\label{sid1}
C (\omega_{\rm sid})={{1}\over{2}}
(1/2 -\beta +\delta) {{ V^2_{\rm sun} }\over{c^2}} |\sin 2\Theta| \sin 2\chi
\EE
and
\BE
\label{sid2}
C (2\omega_{\rm sid})={{1}\over{2}}
(1/2 -\beta +\delta){{V^2_{\rm sun}}\over{c^2}}\cos^2\Theta (1+ \sin^2\chi)
\EE
In the above equations, 
$V_{\rm sun}\sim 369$ km/s and $\Theta\sim -6^o$ indicate the magnitude and
the declination of the solar motion relatively to the CMB while
$ \chi$ is the colatitude of the laboratory 
(for D\"usseldorf $\chi\sim 39^o$).

In this way, one obtains two very different estimates of the RMS parameter.
In fact, on the one hand, the value
$C (\omega_{\rm sid}) \sim (11 \pm 2) \cdot 10^{-16}$ implies
$(1/2 -\beta +\delta) \sim (71 \pm 13)\cdot 10^{-10}$. 
On the other hand, from the analogous result
$C (2\omega_{\rm sid}) \sim (1 \pm 2) \cdot 10^{-16}$, one finds
$(1/2 -\beta +\delta) \sim (1 \pm 2)\cdot 10^{-10}$.

Of course, the 
value 
$(1/2 -\beta +\delta) = (-0.5 \pm 3)\cdot 10^{-10}$ was obtained in 
Ref.\cite{schiller} from a global
fit where also the data for the amplitudes $\hat{B}(t)$ were included.
However, these other data are constrained by the same type of relations 
(see note [20] of Ref.\cite{schiller}) and, therefore, the global fit reflects
the same type of tension between the very different modulations at
$\omega_{\rm sid}$ and $2\omega_{\rm sid}$.

As far as we can see, 
both determinations of $(1/2-\beta +\delta)$ are likely affected
by a systematic uncertainty of theoretical nature. 
In fact, if we consider the relative weight
(for the latitude of D\"usseldorf)
\BE
\label{ratio}
          R\equiv  {{ C (2\omega_{\rm sid}) }\over{
           C(\omega_{\rm sid}) }} \sim 
{{0.7}\over{|\tan \Theta|}} 
\EE
its present experimental value (in February) 
\BE
\label{expdecl}
           R^{\rm EXP}_{\rm feb} \sim 0.09^{+0.18}_{-0.09}
\EE
is very far from
its theoretical prediction for the cosmic motion relatively to the 
CMB, namely 
\BE
\label{ratio1}
          R^{\rm CMB} \sim  6.8
\EE
Therefore, to explain the observed
daily modulations embodied in $C (\omega_{\rm sid})$, one has to consider some
other type of cosmic motion and replace 
the CMB with another possible choice of preferred frame. In this case, 
the experimental determination of the RMS parameter will likely be affected
as well. 

To address the problem from a general point of view, 
let us first return to Eq.(\ref{amplitude}) and introduce the time-dependent
amplitude of the ether-drift effect 
\BE
\label{amplitude1}
A(t)= v^2(t) X 
\EE
in terms of the
Earth's velocity in the plane of the interferometer $v(t)$ and of
the correct unknown normalization of the experiment $X$. 
The main point is that 
the relative variations of the signal
depend only on the kinematic
details of the given cosmic motion and, as such, can be predicted independently
of the knowledge of $X$. To predict the variations of $v(t)$, we shall use
the expressions given by Nassau and Morse \cite{nassau}. These have the
advantage of being fully model-independent and
extremely easy to handle. 
Their simplicity depends on the introduction of a cosmic Earth's velocity 
\BE
\label{vincl}
{\bf{V}}={\bf{V}}_{\rm sun} +
{\bf{v}}_{\rm orb}
\EE 
that, in addition to the genuine cosmic motion of the solar system defined by
${\bf{V}}_{\rm sun}$, includes 
the effect of the Earth's orbital motion around the Sun described by
${\bf{v}}_{\rm orb}$. To a very good approximation, 
${\bf{V}}$ can be taken to be
constant within short observation periods of 2-3 days. Therefore, 
by introducing the latitude
of the laboratory $\phi$, the right
ascension $\tilde{\Phi}$ and the declination $\tilde{\Theta}$ 
associated with the vector
${\bf{V}}$, 
the magnitude of the
Earth's velocity in the plane of the interferometer is defined by the two
equations \cite{nassau}
\BE
\cos z(t)= \sin\tilde{\Theta}\sin \phi + \cos\tilde{\Theta}\cos\phi 
\cos(\lambda)
\EE
and 
\BE
\label{vearth}
v(t)=V \sin z(t)
\EE
$z=z(t)$ being the zenithal distance of ${\bf{V}}$. Here, we have introduced
the time $\lambda\equiv \tau -\tau_o-\tilde{\Phi}$, 
$\tau=\omega_{\rm sid}t$ being the sidereal time of the observation
in degrees and $\tau_o$ being an offset that, in general, 
has to be introduced to compare
with the definition of sidereal time adopted in Ref.\cite{schiller}.

Now, operation of the interferometer provides the minimum and maximum daily
values of the amplitude and, as such, the values
$v_{\rm min}$ and $v_{\rm max}$ 
corresponding to $|\cos(\lambda)|=1$. In this way, using the above relations
one can determine the pair of values
$(\tilde{\Phi}_i,\tilde{\Theta}_i)$, $i=1,2,..n$, for each of the $n$ short
periods of observations taken
during the year, and thus plot the direction of the vectors 
${\bf{V}}_i$ on the celestial sphere. Actually, since the ether-drift is a 
second-harmonic effect in the rotation angle of the interferometer, a single
observation is unable to distinguish the pair 
$(\tilde{\Phi}_i,\tilde{\Theta}_i)$ from the pair
$(\tilde{\Phi}_i +180^o,-\tilde{\Theta}_i)$. 
Only repeating the observations in different 
epochs of the year one can resolve the ambiguity. 
Any meaningful ether-drift, in fact, has to
correspond to pairs 
$(\tilde{\Phi}_i,\tilde{\Theta}_i)$ lying on an `aberration circle', defined by the
Earth's orbital motion, whose center
$(\Phi,\Theta)$ defines the right ascension and the declination
of the genuine cosmic motion of the
solar system associated with 
${\bf{V}}_{\rm sun}$. If such a consistency is found, 
using the triangle law, one can finally determine
the magnitude $|{\bf{V}}_{\rm sun}|$ starting from the known values of
$(\tilde{\Phi}_i,\tilde{\Theta}_i)$, 
$(\Phi,\Theta)$ and the value 
$|{\bf{v}}_{\rm orb}|\sim $ 30 km/s.

We emphasize that the basic pairs of values 
$(\tilde{\Phi}_i,\tilde{\Theta}_i)$ determined in this way
only depend on the {\it relative} magnitude of
the ether drift effect, namely on the ratio
${{ v_{\rm min} }\over{v_{\rm max} }}$, in the various periods. 
As such, they are insensitive to any possible theoretical and/or experimental
uncertainty that can affect multiplicatively the {\it absolute}
normalization of the signal.

For instance, suppose one measures a
relative frequency shift $\delta \nu/\nu={\cal O}(10^{-15})$. 
Assuming a value $(1/2-\beta+\delta)\sim 10\cdot  10^{-10}$ 
in Eq.(\ref{amplitude}), this would be interpreted in terms
of a velocity $v \sim 300$ km/s. 
Within Galileian relativity, where one predicts the same 
expressions by simply replacing 
$(1/2-\beta+\delta) \to 1/2$, 
the same frequency shift would be interpreted in terms of
a velocity $v \sim 14$ m/s. Nevertheless, from
the {\it relative} variations of the ether-drift effect 
one would deduce the same pairs 
$(\tilde{\Phi}_i,\tilde{\Theta}_i)$ and, as 
such, exactly the same type of cosmic motion. 
Just for this reason, Miller's determinations with this method, namely
\cite{miller} 
${\bf{V}}_{\rm sun}\sim 210$ km/s, 
$\Phi \sim 74^o$ and $\Theta \sim -70^o$, should be taken seriously.

We are aware that Miller's observations have been 
considered spurious by the authors of Ref.\cite{shankland} as partly due to
statistical fluctuations and/or thermal fluctuations. 
However, to a closer look (see the discussion given in Ref.\cite{cimento})
the arguments
of Ref.\cite{shankland} are not so solid as they appear by reading the abstract
of that paper. Moreover, Miller's solution is {\it doubly}
internally consistent since
the aberration circle due to the Earth's orbital motion was obtained 
in two different and independent ways
(see Fig. 23 of Ref.\cite{miller}). In fact, 
one can determine the basic pairs
$(\tilde{\Phi}_i,\tilde{\Theta}_i)$ either using
the daily variations of the magnitude of the ether-drift effect 
or using
the daily variations of its apparent direction $\theta_0(t)$ 
(the `azimuth') defined, in terms of Eq.(\ref{basic2}), through the relation
$\theta_0(t)=1/2\tan^{-1}({{\hat{B}(t)}\over{\hat{C}(t)}})$. Since 
the two methods were found to give consistent results, 
in addition to the standard choice of preferred frame represented by the
CMB, it might be worth to consider the predictions associated with 
the cosmic motion deduced by Miller. 

Replacing Eq.(\ref{vearth}) into Eq.(\ref{amplitude}) and
adopting a notation of the type
introduced in Ref.\cite{mewes}, we can
express the theoretical amplitude of the signal as 
\BE
\label{amorse}
        {{A(t)}\over{\nu}}=A_0 + 
A_1\sin\tau +A_2 \cos\tau + A_3\sin(2\tau) +A_4 \cos(2\tau)
\EE 
where
\BE
\label{a0}
A_0=(1/2-\beta+\delta)
{{ V^2 }\over{c^2}} (1- \sin^2\tilde{\Theta}\cos^2\chi - {{1}\over{2}}
\cos^2\tilde{\Theta}\sin^2\chi)
\EE
\BE
\label{a1}
A_1=-{{1}\over{2}}(1/2-\beta+\delta)
{{ V^2 }\over{c^2}} \sin 2\tilde{\Theta} 
\sin(\tilde{\Phi} +\tau_o) \sin 2\chi
\EE
\BE
\label{a2}
A_2=-{{1}\over{2}}(1/2-\beta+\delta)
{{ V^2 }\over{c^2}} \sin 2\tilde{\Theta} 
\cos(\tilde{\Phi} +\tau_o) \sin 2\chi
\EE
\BE
\label{a3}
A_3=-{{1}\over{2}}(1/2-\beta+\delta)
{{ V^2 }\over{c^2}} \cos^2\tilde{\Theta} 
\sin[2(\tilde{\Phi} +\tau_o)] \sin^2\chi
\EE
\BE
\label{a4}
A_4=-{{1}\over{2}}(1/2-\beta+\delta)
{{ V^2 }\over{c^2}} \cos^2\tilde{\Theta} 
\cos[2(\tilde{\Phi} +\tau_o)] \sin^2\chi
\EE
Recall that $V$, $\tilde{\Theta}$ and $\tilde{\Phi}$ 
indicate respectively
  the magnitude, the declination and the right ascension
  of the velocity defined in Eq.(\ref{vincl}). As such, they change during the
year. Also, Eqs.(\ref{a0})-(\ref{a4}) enter the {\it full} amplitude of the 
signal $A=\nu\sqrt{ \hat{C}^2 + \hat{B}^2 }$. 
Therefore, it is not so simple to express 
the coefficients $A_0,A_1,A_2,A_3,A_4$ in terms of
the analogous coefficients entering $\hat{C}(\tau)$ and $\hat{B}(\tau)$.

As explained above, with an appropriate data taking in different epochs of
the year, Eqs.(\ref{amorse})-(\ref{a4}) can be used to deduce the basic 
parameters of the Earth's cosmic motion from the daily variations of the
measured frequency shifts. Here we shall follow the other way around and 
explore the implications of Miller's cosmic solution for the experiment
of Ref.\cite{schiller}. To this end, 
we shall start from observations performed
around February 6th-8th using the entries reported in
Tables I and II of Ref.\cite{miller}. In this case, by
restricting to the southern apex pairs 
$\tilde{\Phi}_{\rm feb}\sim 90^o$ and
$\tilde{\Theta}_{\rm feb}\sim -77^o$, 
for the latitude of D\"usseldorf $\phi \sim 51^o$, one predicts a minimum
velocity
$v_{\rm min} \sim 0.44~ V_{\rm feb}$ and a maximum velocity
$v_{\rm max} \sim 0.79~ V_{\rm feb}$. Therefore, 
introducing the unknown normalization in February, say
$X_{\rm feb}$, such that
    $A_{\rm min} \sim (0.44)^2~X_{\rm feb}$ and
    $A_{\rm max} \sim (0.79)^2~X_{\rm feb}$, we obtain
a mean theoretical value 
\BE
\label{av1}
\langle A\rangle\equiv {{1}\over{2}}(
A_{\rm min} +
    A_{\rm max} ) \sim 0.41~
X_{\rm feb}
\EE
and a daily modulation
\BE
\label{feb}
          (\Delta A)_{\rm feb} = \pm (
\langle A\rangle-
A_{\rm min} ) \sim \pm 0.22 X_{\rm feb}
\EE
In this way, if one could 
subtract out from the data of Ref.\cite{schiller} the spurious systematic
component and obtain the true
experimental signal
$\langle A^{\rm exp}\rangle_{\rm true}$ in terms of the correct 
normalization $X_{\rm feb}$, the above relations 
amount to predict a daily modulation
\BE
\label{feb2}
          (\Delta A)_{\rm feb} 
\sim \pm 0.53 \langle A^{\rm exp}\rangle_{\rm true}
\EE
By repeating the same analysis for observations performed around 
September
15th, where the relevant values found by Miller were
$\tilde{\Phi}_{\rm sept}\sim 75^o$ and
$\tilde{\Theta}_{\rm sept}\sim -62^o$ one predicts, again for the latitude of 
D\"usseldorf, 
$v_{\rm min} \sim 0.19~ V_{\rm sept}$ and a maximum velocity
$v_{\rm max} \sim 0.92~ V_{\rm sept}$. Therefore, in terms of the
arbitrary normalization in September, one finds
    $A_{\rm min} \sim (0.19)^2~X_{\rm sept}$ and
    $A_{\rm max} \sim (0.92)^2~X_{\rm sept}$, with 
a mean theoretical value 
\BE
\label{av2}
\langle A\rangle\equiv {{1}\over{2}}(
A_{\rm min} +
    A_{\rm max} ) \sim 0.44~X_{\rm sept}
\EE
and the considerably larger daily modulation
\BE
\label{sept}
          (\Delta A)_{\rm sept} \sim \pm 0.91
\langle A^{\rm exp}\rangle_{\rm true}
\EE
This represents a $\sim +70\%$ increase \cite{consoli}
with respect to the February value in Eq.(\ref{feb2}).
In this way, neglecting the small modulation at $2\omega_{\rm sid}$, 
and comparing Eqs.(\ref{feb}) and (\ref{sept}), one 
predicts an increase of the parameter associated with the daily modulation
$C (\omega_{\rm sid}) \sim (19 \pm 2) \cdot 10^{-16}$ around
September 15th, starting from its February value
$C (\omega_{\rm sid}) \sim (11 \pm 2) \cdot 10^{-16}$ (within the present
normalization of the experiment).

Here, we are assuming that the {\it central} value of the ether-drift effect, 
namely the quantity
$\langle A^{\rm exp}\rangle_{\rm true}$, does not change too much during the
year. This assumption is motivated by the modest difference between the
average values in Eqs.(\ref{av1}) and (\ref{av2}). It 
is also consistent with the re-analysis of Miller's data
performed in Ref.\cite{shankland} where it was found that the 
average magnitudes of the second-harmonic components were only slightly
changing from one epoch to the other (see page 170 of Ref.\cite{shankland}).

Let us now consider the equivalent of the 
relative weight defined in Eq.(\ref{ratio})
\BE
\label{tilderatio}
          \tilde{R}\equiv  {{ A (2\omega_{\rm sid}) }\over{
           A(\omega_{\rm sid}) }} 
\EE
where
\BE
\label{asid}
           A (\omega_{\rm sid})\equiv \sqrt{A^2_1 + A^2_2}
\EE
and
\BE
\label{a2sid}
           A (2\omega_{\rm sid})\equiv \sqrt{A^2_3 + A^2_4}
\EE
Although $\tilde{R}$ is not immediately readable from the 
numbers reported in Ref.\cite{schiller}, its estimate through
the approximate relation (for the latitude
of D\"usseldorf) 
\BE
\label{tildeapp}
          \tilde{R}\sim  {{0.2}\over{|\tan \tilde{\Theta} |}}
\EE
shows that for Miller's solution the weight of the modulation at 
$2\omega_{\rm sid}$ in the overall daily change remains small. 
In fact, $\tilde{R}$ 
evolves from $\sim 0.05$ to $\sim 0.11$ when $\tilde{\Theta}$ changes from
$\tilde{\Theta}\sim -77^o$ in February to 
$\tilde{\Theta}\sim -62^o$ in September. By comparing with 
Eq.(\ref{ratio}), this confirms that
replacing the CMB with another cosmic solution that exhibits 
$|\Theta|\sim 70^o$, one can obtain a small value of 
\BE
R\sim 3.5 \tilde{R}\sim 0.25
\EE
in Eq.(\ref{ratio}) and thus
consistent estimates
of $(1/2 -\beta +\delta)$ from the two sets of observables at 
$\omega_{\rm sid}$ and $2\omega_{\rm sid}$. 

\section{Summary and outlook}

In this paper we have explored some phenomenological consequences of assuming
the existence of a preferred frame. This scenario, that on the one hand
leads us back to the old Lorentzian version of relativity, is on the other 
hand favoured by present models with extra space-time dimensions where
the interactions with the gravitons change
the vacuum into a physical medium with a non-trivial refractive index. 

Our point is that, besides the effect of graviton loops considered so far, 
one should also take into account the existence of the
{\it background} gravitational fields. In fact, they
produce exactly the same effect transforming the
local vacuum into a physical medium 
whose refractive index can be 
easily computed from the weak-field isotropic form of the metric. Thus, 
if there were a preferred frame $\Sigma$ where light is seen isotropic, 
one should be able 
to detect some effect with the new generation of
precise ether-drift experiments using
rotating cryogenic optical resonators. In particular, 
one should look for
periodic modulations of the signal
that might be associated with the Earth's rotation 
and its orbital motion around the Sun.

When comparing with the experimental results of Ref.\cite{schiller} 
(obtained around February 6th, 2005) we can
draw the following conclusions. The data exhibit a clear modulation 
at the Earth's rotation frequency embodied in the coefficient 
\BE
           C (\omega_{\rm sid}) \sim (11 \pm 2) \cdot 10^{-16}
\EE
that might represent the first modern experimental evidence for 
a preferred frame. At the same time, the signal at $2\omega_{\rm sid}$ 
\BE
           C (2\omega_{\rm sid}) \sim (1 \pm 2) \cdot 10^{-16}
\EE
is very weak. Thus, the experimental value of the ratio
\BE
\label{present}
         R^{\rm EXP}_{\rm feb}\equiv {{ C (2\omega_{\rm sid}) }\over{
           C(\omega_{\rm sid}) }} \sim 0.09^{+0.18}_{-0.09}
\EE
is very far from the expected theoretical value for the Earth's motion
relatively to the CMB
\BE
            R^{\rm CMB} \sim 6.8
\EE
For this reason, to explain the daily modulations, one should consider some
other type of cosmic motion. As an example, we have explored the implications 
of the cosmic motion 
deduced from Miller's ether-drift observations. In this framework, 
one expects a modest
daily modulation at $2\omega_{\rm sid}$ in all periods of the
year. This prediction is 
in good agreement with the present experimental value Eq.(\ref{present})
and will be tested with future measurements.

Within Miller's cosmic solution, one also predicts a $\sim +70\%$ 
increase of the daily modulation, from its February value 
$C (\omega_{\rm sid}) \sim (11 \pm 2) \cdot 10^{-16}$ up to 
$C (\omega_{\rm sid}) \sim (19 \pm 2) \cdot 10^{-16}$ in September
(within the present normalization of the experiment).

This other prediction will also
be tested with experimental data collected
in the next few months and, whenever confirmed, would represent clean 
experimental evidence for the existence of a preferred frame, 
a result with far-reaching implications for both
particle physics and cosmology.
\vskip 20 pt
\centerline{\bf{Acknowledgements}}
M. C. thanks N. Cabibbo and P. M. Stevenson for 
useful discussions. C.M.L. de A. thanks the Italian Ministero degli
Affari Esteri (MAE) for financial support.

\end{document}